# How much non-coding DNA do eukaryotes require?

Sebastian E. Ahnert[1,3], Thomas M. A. Fink[2] & Andrei Zinovyev[1]

[1] *Institut Curie, Bioinformatics, 26 rue d'Ulm, Paris 75248, France*
[2] *Institut Curie, CNRS UMR 144, 26 rue d'Ulm, Paris 75248, France*
[3] *Theory of Condensed Matter, Cavendish Laboratory, University of Cambridge, Cambridge CB3 0HE, UK*

**Despite tremendous advances in the field of genomics, the amount and function of the large non-coding part of the genome in higher organisms remains poorly understood. Here we report an observation, made for 37 fully sequenced eukaryotic genomes, which indicates that eukaryotes require a certain minimum amount of non-coding DNA (ncDNA). This minimum increases quadratically with the amount of DNA located in exons. Based on a simple model of the growth of regulatory networks, we derive a theoretical prediction of the required quantity of ncDNA and find it to be in excellent agreement with the data. The amount of additional ncDNA (in basepairs) which eukaryotes require obeys $N_{DEF} = 1/2 \, (N_C / N_P) (N_C - N_P)$, where $N_C$ is the amount of exonic DNA, and $N_P$ is a constant of about 10Mb. This value $N_{DEF}$ corresponds to a few percent of the genome in Homo sapiens and other mammals, and up to half the genome in simpler eukaryotes. Thus our findings confirm that eukaryotic life depends on a substantial fraction of ncDNA and also make a prediction of the size of this fraction, which matches the data closely.**

In most eukaryotes, a large proportion of the genome does not code for proteins. The non-coding part of eukaryotic genomes has likely been expanded by various mechanisms throughout evolution, such as insertions and deletions of DNA sequence segments[1] and whole genome duplication[2]. Unlike the coding part, it is observed to vary greatly in size even between closely related species[3,4]. Several recent large-scale efforts to catalogue genome sizes, for example the Animal Genome Size Database[5], Plant DNA C-values database[6] and Fungal Genome Size Database[7], have provided striking new examples of this (see Appendix A). There now exists an accumulation of evidence that non-coding DNA (ncDNA) in eukaryotes is genetically active, and that it is likely to play an important role in genetic regulation[8,9,10]. In particular, very short, specific sequences of ncDNA have been discovered, which give rise to non-coding RNAs (ncRNAs) such as microRNA[11-14] and siRNA[15,16]. These RNAs are known to fulfill regulatory functions and have also been linked to diseases in humans[17]. Since systematic efforts to catalogue ncRNA sequences have only begun recently[18], it is likely that many more remain to be discovered. More circumstantial but equally intriguing evidence is the conservation of non-coding sequences in mammals revealed by genome comparisons[19].

In this paper we analyze the total amounts of protein-coding-related DNA (which we shall term $N_C$) and ncDNA ($N_{NC}$) in 330 prokaryotes and 37 eukaryotes. All prokaryotes and eukaryotes have been fully sequenced, and the fraction of protein-coding-related DNA for these species can be found in databases of genome statistics[20]. In eukaryotes we take $N_C$ to be the total length of all exons in the genome. This includes UTRs which are essential to the protein-coding machinery.

$N_C$ and $N_{NC}$ values can be considered as independent variables, which we plot against each other in Figure 1. This plot reveals the following facts:

i. In prokaryotes the total length of ncDNA increases linearly with the total length of protein-coding DNA.
ii. Prokaryotes and eukaryotes form two separate regimes in the space of protein-coding-related DNA and ncDNA lengths, which meet around $N_C = 10^7$ and $N_{NC} = 10^6$.
iii. In eukaryotes the total length of ncDNA is bounded from below by a quadratic function of the total length of exonic DNA.

Our first conclusion, based on 330 prokaryotic genomes (all available in GenBank in March, 2007), confirms recent findings that a constant fixed non-coding proportion (~12%) of a prokaryotic genome is required for organism's functioning:[21] presumably, this is space occupied by promoter regions and by infrastructural RNAs required for protein synthesis. Our analysis did not show any significant difference of this proportion between bacterial and archaea genomes, or any relation with general genome composition features such as GC-content.

The second conclusion demonstrates the 'complexity ceiling' of prokaryotic organisms[9,22,23], whose regulation is based almost solely on proteins. It however also contains an intriguing observation about the continuity of the transition between the prokaryote and eukaryote worlds: the simplest eukaryotes have $N_{NC}$ and $N_C$ values close to those of the most complex bacteria. It is not obvious that this should be so, due to the great structural changes in eukaryote cells in comparison to prokaryotes, such as the cell nucleus and chromatin, the appearance of introns and splicing, and the organization of genome into chromosomes. These features are found in all eukaryotes, but are absent from almost all prokaryotes. The only exceptions are some prokaryotes which have evolved a simplified versions of chromatin[24] or introns[25]. Given the time scale of genome evolution, it is surprising that $N_{NC}$ values in the simplest eukaryotes have not significantly diverged from the small values found in prokaryotes. This suggests that the tremendous reorganization of the eukaryotic cell did not substantially alter the factors determining the required fraction of ncDNA.

Our third conclusion is the most important. It connects the size of the exonic part of the eukaryotic genome ($N_C$) to a lower bound on the size of its non-coding part ($N_{NC}$), the latter scaling quadratically with the former. In prokaryotes, a quadratic relationship between the number of regulatory genes and the total gene number in prokaryotes has been demonstrated in the literature[26]. Unlike prokaryotes, who encode all their regulatory overhead in genes, eukaryotes are able to recruit non-coding DNA for this purpose[9,21,27]. Figure 1 suggests that the quadratic relationship between regulatory overhead and protein-coding DNA observed in prokaryotes still holds for eukaryotic genomes in the form of a lower bound on $N_{NC}$, which represents the amount of ncDNA required to encode additional regulatory information. Based on this hypothesis we derive a theoretical prediction for the lower bound on the amount of eukaryotic ncDNA by considering a simple accelerating growth model of regulatory genetic networks[28]. Despite the simplicity of our model, this prediction matches the data closely (see Figure 1 and text below).

In a regulatory genetic network we take the nodes to be protein-coding genes, and edges to represent gene regulations. The number of nodes $c$ and the number of edges $n$ are related by the average degree $k = 2n/c$ (the average number of connections per node).

Our model rests on two conditions. The first is that in protein interaction networks the capacity of one node to connect to other nodes is limited.[28] This is a constraint on any physical or biological network which is evident, for example, in the limited connectivity in computer chip architecture or the urban planning of roads. Similarly, recent research has revealed that the number of protein interfaces required for regulating other proteins and receiving regulation in yeast is limited to 14.[29] In general this means that $k$ cannot exceed some maximum value $k_{max}$.

The second condition is that protein regulatory networks are highly integrated systems which rely on global connectivity. In regulatory networks the average node is connected to a fixed fraction of the other nodes, even though this fraction may be much less than 1; such networks are called *accelerating networks*[9,26,28,30]. In such networks the total number of connections is proportional to the square of the number of genes. In the language of our model, $n = \alpha c^2$, where $\alpha$ is a constant of proportionality. Intuitively, introducing a new function in the system requires adding new genes that should be regulated. However, a fraction of these regulators also needs additional regulation to integrate the new function in the system (for example, for feedback function control), which results in faster-than-linear growth of the number of regulations[26].

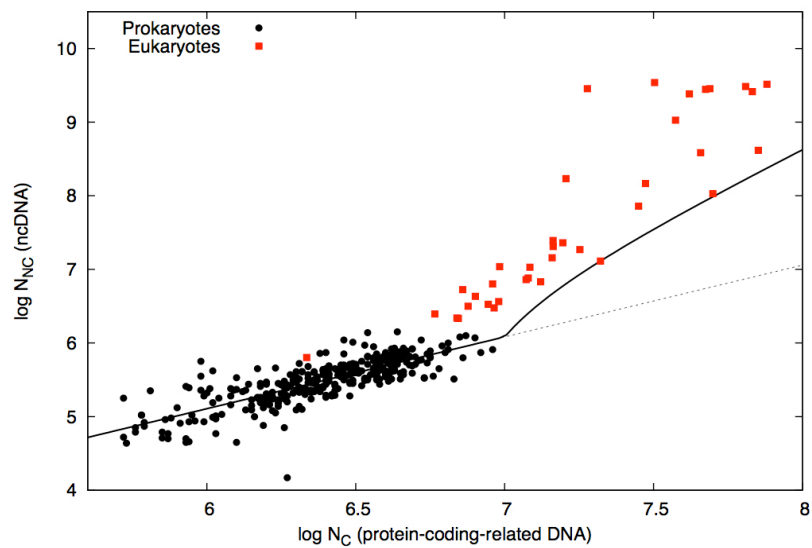

**Figure 1** Values of protein-coding-related ($N_C$) versus non-coding ($N_{NC}$) DNA for 330 prokaryotic (black circles) and 37 eukaryotic (red squares) species in units of basepairs, on a logarithmic scale. The prokaryotes exhibit linear growth of $N_{NC}$ with $N_C$, as has been suggested in the literature[21,22]. A best fit $N_{NC} = 0.181 \, N_C^{0.975}$ (dotted line, identical to solid line below $N_C = 10^7$) provides strong evidence of this linear relationship. For the eukaryotes on the other hand, $N_{NC}$ grows much more rapidly with $N_C$, and is bounded from below. Thus, while eukaryotes can have almost arbitrarily large amounts of non-coding DNA, there appears to be a necessary minimum amount which is required, depending on the amount of exonic DNA. The solid line shows the theoretical lower limit $N_{MIN}$ on the amount of non-coding DNA in eukaryotes as derived in the text and given by equation (3). Note that there is a eukaryote (*E. cuniculi*) with a very short genome among the prokaryotes. The entire eukaryotic dataset, as well as a description of how the data points were collated and derived, can be found in the supplementary information (Table S1).

Due to the second condition, the average degree $k$ grows with the network size $c$. For a sufficiently large network, the maximum connectivity $k_{max}$ is reached and it is no longer possible to maintain global integration. This defines a critical network size $c_{crit}$, above which the two conditions are in conflict. The first states that $k_{max} > 2 n / c$. Combining this with the second one allows us to write $c < k_{max} / 2 \alpha$. Thus, as the number of genes $c$ grows, it encounters the threshold $c_{crit} = k_{max} / 2 \alpha$. We can thus rewrite the constant $\alpha$ as $\alpha = k_{max} / 2 c_{crit}$. Were $c$ to grow beyond this threshold, a total of $n = \alpha c^2 = c^2 k_{max} / 2 c_{crit}$ connections would be required, but only $n_p = c k_{max} / 2$ would be present, giving rise to a deficit of connections $n_{def}$:

$$n_{def} = n - n_p = (k_{max} / 2) (c / c_{crit}) (c - c_{crit}) \qquad (1)$$

See Figure 2 for an illustration of this derivation.

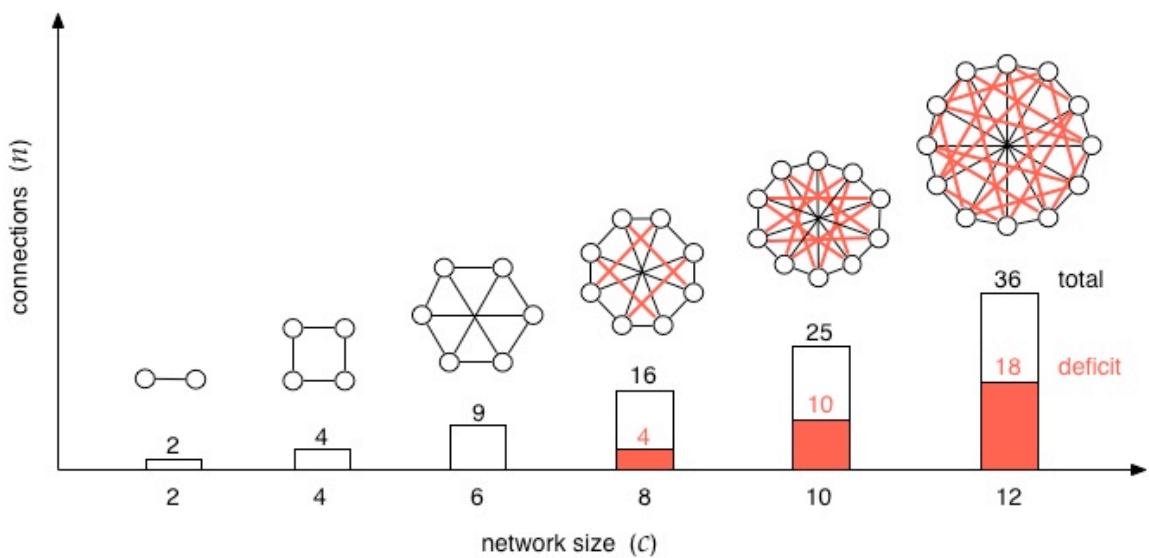

**Figure 2** Deficit in regulatory networks as given by equation (1): In regulatory networks the need for integrated connectivity means that the number of connections (gene regulations) $n$ scales quadratically with the number of nodes (genes) $c$. This is equivalent to saying that the average degree $k$ grows linearly with $c$. In the simple networks in this figure, every node is connected to half of all the nodes so that $k = c / 2$. If, however, $k$ cannot exceed a given value $k_{max}$ (here equal to 3), then the network accrues a deficit of connections (drawn in red), $n_{def} = (3 / 2)(c / 6)(c - 6) = c^2 / 4 - 9 c$ when it grows beyond $c_{max} = 2 k_{max} = 6$. The bar below each network shows the deficit (red) as a proportion of the total.

In our model we associate $c_{crit}$ with the limiting size of a regulatory network based solely on the regulation of proteins by other proteins. Our hypothesis is that $c_{crit}$ divides life into prokaryotes ($c < c_{crit}$) and eukaryotes ($c > c_{crit}$).

We assume that the genetic network must be encoded in the genome in some form. Let us say that the nodes of the network (representing proteins) require $N_C = l_c c$ nucleotides, where $l_c$ is the average mRNA length. In prokaryotes, almost all regulations are performed through protein-protein interfaces (encoded in $N_C$) and protein-DNA interfaces (encoded in a relatively small $N_{NC}$). In eukaryotes, additional regulatory connections (represented by network edges) are necessary in order to cancel the deficit $n_{def}$ and these connections are encoded in $N_{NC}$[9,21,27]. Let us assume that the average cost of encoding one such additional regulation is $l_n$. Hence, the total cost of additional deficit regulations is $N_{DEF} = l_n n_{def}$. Then we find:

$$N_{DEF} = (\beta / 2) (N_C / N_{CC}) (N_C - N_{CC}) \qquad (2)$$

where $\beta = k_{max} l_n / l_c$ is a constant, and $N_{CC} = l_c c_{crit}$ is the maximum size of the protein-coding part of the prokaryotic genome, which is known to be approximately 10Mb[31].

One last step is required before we can compare the prediction of equation (2) to the data. We have fitted the data for prokaryotes to a function of the form $N_P = a N_C^b$ for which the parameters are found to be

log($a$) = -0.743 ± 0.257 and $b$ = 0.975 ± 0.040 (Figure 2), which, as $b$ is close to one, implies a linear relationship between $N_{NC}$ and $N_C$ in prokaryotes. As discussed earlier, this confirms suggestions in the literature that the amount of ncDNA in prokaryotes is a roughly constant proportion (~12%) of the total genome, comprising e.g. promoter sequences and infrastructural RNAs required for protein synthesis.[21,22] As these non-coding sequences are also required in eukaryotes, we add $N_{DEF}$ to $N_P$ in order to obtain the prediction $N_{MIN}$ of the total required amount of ncDNA for $N_C > N_{CC}$:

$$N_{MIN} = N_P + N_{DEF} = 0.181\, N_C^{0.975} + (\beta / 2)\, (N_C / N_{CC})\, (N_C - N_{CC}) \quad (3)$$

In Figure 1 we show $N_P$ and the prediction for $N_{MIN}$. This prediction forms a lower bound on the amount of ncDNA which closely matches the boundary of the data points across the entire range of eukaryotic species. As $N_C$ increases beyond $N_{CC}$, the gap between $N_{MIN}$ and $N_P$ widens, representing the additional ncDNA required to balance the deficit of regulatory connections $N_{DEF}$.

**Table 1 Genome statistics for 9 eukaryotic species**

| Species | L | $N_{NC}$ | $N_C$ (% of L) | $N_{DEF}$ (% of L) | $N_{MIN}$ (% of L) |
|---|---|---|---|---|---|
| Homo sapiens | 3107 | 3043 | 64 (2.1) | 159 (5.1) | 167 (5.4) |
| Rattus norvegicus | 2834 | 2787 | 47 (1.7) | 80 (2.8) | 86 (3.0) |
| Mus musculus | 2664 | 2597 | 68 (2.5) | 179 (6.7) | 186 (7.0) |
| Gallus gallus | 1100 | 1063 | 37 (3.4) | 47 (4.3) | 51 (4.7) |
| Oryza sativa | 430 | 384 | 45 (10.6) | 73 (17.1) | 79 (18.3) |
| Drosophila melanogaster | 176 | 147 | 30 (16.8) | 27 (15.1) | 30 (17.1) |
| Caenorhabiditis elegans | 100 | 72 | 28 (28.1) | 23 (23.2) | 27 (26.5) |
| Dictyostellum discoideum | 34 | 13 | 21 (61.8) | 10 (30.8) | 13 (38.2) |
| Plasmodium falciparum | 23 | 11 | 12 (53.2) | 1 (5.2) | 3 (11.6) |

Genome length $L$, amounts of non-coding DNA $N_{NC}$ and exonic DNA $N_C$, predicted regulatory ncDNA $N_{DEF}$ and total minimum of ncDNA $N_{MIN} = N_{DEF} + N_P$ for $\beta$ = 0.909. $L$, $N_{NC}$, $N_C$ and $N_{MIN}$ are given in millions of basepairs. The values in brackets are $N_C$ and $N_{MIN}$ as percentages of $L$. The values of $N_{NC}$ and $N_C$ are derived from exon statistics of the NCBI database[20].

The only free parameter in equation (3) is $\beta$, which is fitted to the data (see Appendix B for details). This gives a value of $\beta$ = 0.909. In fact, a compelling argument can be made for why $\beta$ should be equal to 1 (see Appendix C). The predicted and observed fact of that $\beta$ is close to one means that $l_n \approx l_c / k_{max}$. This in turn implies that if $k_{max}$ (maximum number of protein interfaces to other proteins) is much larger than one, ($k_{max} >> 1$) then $l_n << l_c$. As a result the new regulatory system provided by the 'deficit' network edges is significantly cheaper than the purely protein-based prokaryotic regulations, where the cost is measured in nucleotides per regulation. This is also in agreement with predictions in the literature[8,9,27], where non-coding RNA-based regulation is compared to cheaper 'digital' components of the system as opposed to the more costly 'analogue' protein-protein regulation.

In our network model ncRNAs are represented as deficit edges, which are cheap connectors between proteins. However, no assumptions are made about the likely mechanism of such interactions. It is unlikely that small ncRNAs can serve directly as adaptors for protein-protein interactions. However, the fact that small RNA molecules only require a short sequence on a protein or mRNA for regulation makes them cheap and easily reconfigurable[23]. For example, if it is advantageous for a transcription factor A to regulate a number of proteins, protein-protein regulation – the 'analogue' architecture - would require developing a new binding sequence in promoter regions of the genes of all these proteins, or developing a new protein-protein interface in each target protein. The first way would require redesigning already highly tuned transcriptional machinery for a number of genes, while the second would necessitate the development of new domains in proteins and the design of protein structures to adopt them. By contrast, noncoding RNA-based regulation - the 'digital' architecture - would need much less, and more flexible, intervention: In the above example, it could provide a solution by creating a microRNA regulated by A and its short target sequences in the mRNAs of target proteins. This task is relatively easy because it does not require re-designing existing machinery such as transcription complexes or protein structures and does not result in intensive cross-talk with other system components[23].

Using the formula (3) and the known parameter $\beta$, it is now possible to predict the difference between the observed quantity of ncDNA $N_{NC}$ and required to cover the regulatory deficit $N_{DEF}$. Table 1 lists the values of the predicted amount of regulatory ncDNA ($N_{DEF}$) and the resulting total minimum amount of ncDNA ($N_{MIN}$) for 9 eukaryotic species (see supplementary information for details of all 37 eukaryotes as well as additional data and information on other species). For mammals these values lie around 2-6% of the genome, while for simpler eukaryotes this fraction can be an order of magnitude larger (e.g., over 30% in *Dictyostellum discoideum*).

It is interesting to compare our prediction $N_{MIN}$ of the minimum amount of ncDNA with the amount of conserved ncDNA reported in several known genomes. Although sequence conservation does not determine sequence functionality[32,33], this number gives a reasonable estimate for the lower bound of the functional portion of ncDNA.

For the human genome our prediction of the minimum non-coding amount $N_{MIN}$ is 5.4% of the total genome length (see Table 1). This value is comparable in magnitude with the level of ncDNA conservation between mouse and human genomes, estimated in 3-4%[19].

For the species *Drosophila melanogaster* our prediction gives 17.1% for $N_{MIN}$ as a fraction of L. Recently 12 *Drosophila* species have been sequenced and annotated[34,35]. A careful analysis of sequence conservation, taking into account possible sequence shuffling and flips, revealed that about 12-13% of the total genome of *Drosophila melanogaster* is conserved in intronic and intergenic regions when compared with genomes of the phylogenetically most distant species *D. grimshawi*, *D. virilis* and *D. mojavensis*[36].

Our predictions on the minimum required amount of ncDNA are a few percent higher than the conserved ncDNA in these two well-studied organisms, which provides support for the hypothesis that the importance of ncDNA (including poorly conserved regions) is currently underestimated[27].

## DISCUSSION

The network growth model we construct is a gross simplification of the real evolution of regulatory genetic networks. For example, one implicit approximation used in the model is that the proteins (nodes) in the network remain of the same structural complexity: *i.e.*, their maximal number of interfaces $k_{max}$ and the encoding cost $l_c$ remain the same in all organisms. In reality of course, the regulation based on proteins also evolves from bacteria to human, allowing much denser protein network. Fortunately, in the final formula (3) these factors are cancelled: the only requirement is that the $\beta$ coefficient must be independent on $N_C$. If the cost of encoding deficit edges $l_n$ is constant, then only the proportion $k_{max} / l_c$ should be fixed – in other words the model allows longer proteins in higher organisms to provide more possibilities to connect to other proteins.

It has been suggested, by Mattick[8,9,10] amongst others, that eukaryotes may have evolved from prokaryotes by enlisting substantial amounts of ncRNA for regulatory tasks. It has even been proposed that an entire 'parallel regulatory system' based on ncRNA may lie hidden, with the recently discovered microRNA and siRNA being the tip of the iceberg[27]. It was also speculated that probably as much as 20% of human genome is transcribed in functional ncRNAs[27]. Our predictions in Table 1 do not contradict to this statement, since we estimate only the lower boundary for this number.

The relationships between protein-coding-related DNA and ncDNA revealed by our comprehensive survey, together with the prediction derived from our theoretical model, show (a) that a minimum amount of ncDNA is required in eukaryotes which (b) scales with the number of regulatory connections required for a fully integrated network. This suggests that this fraction of ncDNA codes for regulatory mechanisms – most likely mediated by ncRNA – and thus that large-scale involvement of ncRNA in genetic regulation is likely, providing compelling support for the above hypothesis of the cheaper 'parallel regulatory system'. If this theory of parallel regulation is borne out, then it would seem that systematic large-scale efforts to identify the regulatory role of ncRNA are very likely to be fruitful.

## ACKNOWLEDGEMENTS

We are grateful to the Research Department of Institut Curie (Paris) for providing support and thank Emmanuel Barillot, François Radvanyi and Mike Payne for support and stimulating discussions. We thank the anonymous referee for his or her useful remarks and proposals, which significantly improved the paper. During this work S.E.A. was supported by the Association pour le Recherche sur le Cancer (ARC), France, and The Leverhulme Trust, UK.

# REFERENCES

1. Petrov, D.A. Mutational Equilibrium Model of Genome Size Evolution. *Theor. Pop. Biol.* **61**, 533-546 (2002).
2. Kellis, M., Birren, B. W. & Lander, E. S. Proof and evolutionary analysis of ancient genome duplication in the yeast *Saccharomyces cerevisiae*. *Nature* **428**, 617-624 (2004)
3. Gregory, T. R. Synergy between sequence and size in large-scale genomics. *Nat. Rev. Gen.* **6**, 699-708 (2005).
4. Gregory, T.R. & Hebert, P.D.N. The modulation of DNA content: proximate causes and ultimate consequences. *Genome Res.* **9**, 317-324 (1999).
5. Gregory, T.R. Animal Genome Size Database (2005), www.genomesize.com
6. Bennett, M. D. & Leitch, I. J. Plant DNA C-values database (release 3.0, Dec. 2004),
7. Kullman, B., Tamm, H. & Kullman, K. Fungal Genome Size Database (2005) www.zbi.ee/fungal-genomesize www.rbgkew.org.uk/cval/homepage.html
8. Mattick, J. S. Challenging the dogma: the hidden layer of non-protein-coding RNAs in complex organisms. *BioEssays* **25**, 930-939 (2003).
9. Mattick, J. S. RNA regulation: a new genetics? *Nat. Rev. Genet.* **5**, 316-323 (2004).
10. Mattick, J. S. Non-coding RNAs: the architects of eukaryotic complexity. *EMBO Reports* **2**, 986-991 (2001).
11. Ruvkun, G., Glimpses of a Tiny RNA World. *Science* **294**, 797-799 (2001)
12. Lagos-Quintana, M., Rauhut, R., Lendeckel, W. & Tuschl, T., Identification of Novel Genes Coding for Small Expressed RNAs. *Science* **294**, 853-858 (2001)
13. Lau, N. C., Lim, L. P., Weinstein, E. G. & Bartel, D. P, An Abundant Class of Tiny RNAs with Probable Regulatory Roles in *Caenorhabditis elegans*. *Science* **294**, 858-862 (2001)
14. Lee, R. C. & Ambros, V., An Extensive Class of Small RNAs in *Caenorhabditis elegans*. *Science* **294**, 862-864 (2001)
15. Whalley, K., RNA interference: Breakthrough for systematic RNAi. *Nature Reviews Genetics* **7**, 331 (2006)
16. Couzin, J., Small RNAs Make Big Splash. *Science* **298**, 2296-2297 (2002)
17. McManus, M. T. MicroRNAs and cancer. *Semin. Cancer Biol.* **13**, 253-258 (2003).
18. Pang, K. C., Stephen S., Dinger M. E., Engstrom P. G., Lenhard B., Mattick J. S. RNAdb 2.0—an expanded database of mammalian non-coding RNAs. Nucleic Acids Research **35**, D178–D182 (2007)
19. The Mouse Genome Sequencing Consortium. Initial sequencing and comparative analysis of the mouse genome. *Nature* **420**, 520-562 (2002).
20. National Center for Biotechnology Information (NCBI) database http://www.ncbi.nih.gov/genomes/leuks.cgi - total protein-coding proportion determined using total length of all exons.
21. Taft, R. J., Pheasant, M., Mattick, J. S., The relationship between non-protein-coding DNA and eukaryotic complexity. *Bioessays* **29**, 288 (2007)
22. Taft, R. J. & Mattick, J. S. Increasing biological complexity is positively correlated with the relative genome-wide expansion of non-protein-coding DNA. *arXiv Preprint Archive* http://www.arxiv.org/abs/q-bio.GN/0401020 (2003).
23. Mattick, J. S. A new paradigm for developmental biology. The Journal of Experimental Biology **210**, 1526-1547 (2007).
24. Sandman, K. and Reeve, J. N., Archaeal chromatin proteins: different structure but common function? *Curr. Opin. Microbiol.* **8**, 656 (2005).
25. Toro, N., Bacteria and Archaea Group II introns: additional mobile genetic elements in the environment. *Environ. Microbiol.* **5**, 143 (2003).
26. Croft L. J., Lercher M. J., Gagen M. J., Mattick J. S. Is prokaryotic complexity limited by accelerated growth in regulatory overhead? *Genome Biology* **5:**P2 (2003)
27. Pheasant M., Mattick J. S. Raising the estimate of functional human sequences. Genome Res **17**(9): 1245 – 1253 (2007)
28. Mattick, J. S. & Gagen, M. J. Accelerating networks. *Science* **307**, 856-858 (2005).
29. Kim, P. M., Lu, J. L. , Xia, Y., Gerstein, M. B. *Science* **314**, 1938-1941 (2006).
30. Albert, R., Barabasi, A.-L. The statistical mechanics of complex networks. *Rev. Mod. Phys.* **74**, 47 (2002)
31. Fogel, G. B., Collins, C. R., Li, J. & Brunk, C. F. Prokaryotic genome size and SSU rDNA copy number: estimation of microbial relative abundance from a mixed population. *Microbial Ecology* **38**, 93-113 (1999).
32. Pheasant M., Mattick J. S. Raising the estimate of functional human sequences. Genome Res **17**(9): 1245 – 1253 (2007)
33. Pang, K.C., Frith, M.C., and Mattick, J.S. 2006. Rapid evolution of noncoding RNAs: Lack of conservation does not mean lack of function. Trends Genet. **22:** 1–5.
34. see http://rana.lbl.gov/drosophila
35. Gilbert, D.G. DroSpeGe: rapid access database for new Drosophila species genomes. Nucleic Acids Research **35**, D480-D485 (2007)
36. Martignetti L., Caselle M., Jacq B., Herrmann C. DrosOCB: a high resolution map of conserved non coding sequences in Drosophila. 2007. http://arxiv.org/abs/0710.1570

**Appendix A**

Examples of the C-value paradox: The genome of *Drosophila melanogaster* (the fruit fly) is 176Mb long, while that of *Podisma pedestris* (the mountain grasshopper) is 16.5Gb[5] (94 times as large). The genome of *Oryza sativa* (a variety of rice) is 430Mb long, while that of *Triticum aestivum* (wheat) is 16.9 Gb[7] (39 times as large).

**Appendix B**

The free parameter $\beta$ in equation (2) was set by using the lowest non-inferred data point in the range of $N_C > C_{prok}$, i.e. that of *Dictyostelium discoideum* with $\log(N_C) = 7.32$ and $\log(N_{NC}) = 7.11$. Subtracting the prokaryotic gradient gave us $N_{DEF} = N_{NC} - N_P = 10^{7.11} - 0.182 \times 10^{7.32 \times 0.975} = 1.05 \times 10^7$ and thus, with $N_{CC} = 10^7$ being the critical maximum amount of coding DNA in prokaryotes (see text), we obtain $\beta = (2 N_{DEF} N_{CC}) / (N_C (N_C - N_{CC})) = (2 \times 1.05 \times 10^7 \times 10^7) / (10^{7.32} (10^{7.32} - 10^7)) = 0.909$.

**Appendix C**

If we add a single node to a network of $c$ nodes, then the assumption of integrated connectivity, implying $n = \alpha c^2$, requires that we have to increase the number $n$ of connections by $\Delta n = \alpha [(c + 1)^2 - c^2] = \alpha (2c + 1)$. For $c \gg 1$, $\Delta n \simeq 2 \alpha c = k$, so that adding the $c$th node requires $l_c$ basepairs for the gene and $\Delta n\, l_n = 2 \alpha c\, l_n = k\, l_n$ basepairs for the regulatory overhead of this gene. As $k$ grows proportionally to $c$, the regulatory overhead $k\, l_n$ for a new gene will surpass the gene length $l_c$. In qualitative terms this threshold has been suggested as a reason for the prokaryotic genome ceiling[9,26]. Hence the value of $k$ at the equilibrium $l_c = l_n k$ offers itself as a natural choice for $k_{max}$, so that $l_c = l_n k_{max}$ and thus $\beta = 1$.